\documentclass{article}

\usepackage{arxiv}

\usepackage[utf8]{inputenc} 
\usepackage[T1]{fontenc}    
\usepackage{hyperref}       
\usepackage{url}            
\usepackage{booktabs}       
\usepackage{amsfonts}       
\usepackage{nicefrac}       
\usepackage{microtype}      
\usepackage{lipsum}		
\usepackage{setspace}
\usepackage{graphicx}
\usepackage{mhchem}
\usepackage{doi}
\usepackage{amsmath,amssymb,amsthm}
\usepackage[numbers,sort&compress]{natbib}

\title{Phase-Dependent Excitonic Light Harvesting and Photovoltaic  Limits in Monolayer Y$_2$T\lowercase{e}O$_2$ MOenes}

\author{%
\parbox{0.95\linewidth}{\centering
\href{https://orcid.org/0000-0001-5048-0696}{\includegraphics[scale=0.09]{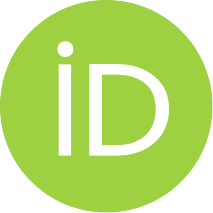}\hspace{1mm}}Bill D. A. Huacarpuma\textsuperscript{1},
\href{https://orcid.org/0000-0002-8366-7227}
{\includegraphics[scale=0.09]{icons/orcid.pdf}\hspace{1mm}}Jos\'e A.~dos S.~Laranjeira\textsuperscript{2},
\href{https://orcid.org/0000-0001-7653-0428}{\includegraphics[scale=0.09]{icons/orcid.pdf}\hspace{1mm}}Nicolas F.~Martins\textsuperscript{2},
\href{https://0000-0002-5217-7145}{\includegraphics[scale=0.09]{icons/orcid.pdf}\hspace{1mm}}Julio R. Sambrano\textsuperscript{2},
\href{https://orcid.org/0000-0003-4699-5886}{\includegraphics[scale=0.09]{icons/orcid.pdf}\hspace{1mm}}Kleuton A. L. Lima\textsuperscript{3},
\href{https://orcid.org/0000-0003-1602-9345}{\includegraphics[scale=0.09]{icons/orcid.pdf}\hspace{1mm}}Santosh K. Tiwari\textsuperscript{4},
\href{https://orcid.org/0000-0001-5934-8528}{\includegraphics[scale=0.09]
{icons/orcid.pdf}\hspace{1mm}}Alexandre C. Dias\textsuperscript{1},
and
\href{https://orcid.org/0000-0001-7468-2946}{\includegraphics[scale=0.09]{icons/orcid.pdf}\hspace{1mm}}Luiz A.~Ribeiro Jr\textsuperscript{1,$\dag$} \\
\vspace{0.6em}
{\normalfont\normalsize
\textsuperscript{1}Computational Materials Laboratory, LCCMat, Institute of Physics, University of Bras\'ilia, 70910-900, Bras\'ilia, Federal District, Brazil\\
\textsuperscript{2}Modeling and Molecular Simulation Group, S\~ao Paulo State University (UNESP), School of Sciences, Bauru 17033-360, SP, Brazil\\
\textsuperscript{3}Department of Applied Physics and Center for Computational Engineering and Sciences, State University of Campinas, Campinas, 13083-859, SP, Brazil\\
\textsuperscript{4}Centre for New Materials and Surface Engineering, Department of Chemistry, NMAM Institute of Technology (NMAMIT), Nitte (Deemed to be University), Nitte 574110, Karnataka, India\\
\vspace{0.6em}
\href{https://scholar.google.com/citations?user=oVFICZQAAAAJ&hl=en}{\includegraphics[scale=0.05]{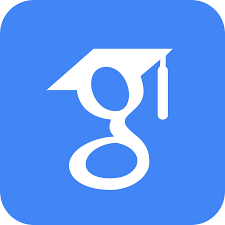}} \href{https://www.linkedin.com/in/bill-darwin-aparicio-huacarpuma-316060255/}{\includegraphics[scale=0.05]{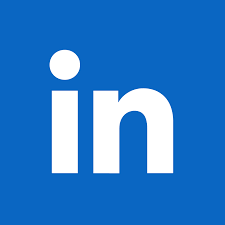}}\hspace{0.1cm}\texttt{\textsuperscript{*}bdaparicioh@gmail.com } \\
\vspace{0.1cm}
\href{https://scholar.google.com/citations?user=EgsxcaUAAAAJ\&hl=pt-BR}{\includegraphics[scale=0.05]{icons/gscholar.png}} \href{www.linkedin.com/in/luiz-ribeiro-164221225}{\includegraphics[scale=0.05]{icons/linkedin.png}}\hspace{0.1cm}\texttt{\textsuperscript{$\dag$}ribeirojr@unb.br}\\
}
}%
}

\renewcommand{\shorttitle}{Bill D.~A.~Huacarpuma et al.}

\hypersetup{
pdftitle={3d-dodeca},
pdfsubject={3d-dodeca},
pdfauthor={Bill D.~A.~Huacarpuma, Luiz A.~Ribeiro Jr},
pdfkeywords={MOenes, Y$_2$TeO$_2$ monolayers, excitonic effects, Bethe--Salpeter equation, photovoltaic efficiency.},
}

\begin{document}
\maketitle

\onehalfspacing

\begin{abstract}
We investigate phase-dependent electronic and excitonic phenomena in monolayer \ce{Y2TeO2} MOenes in the 1T and 2H polymorphs using first-principles theory and an effective many-body framework. Phonon spectra and elastic stability criteria establish both phases as dynamically and mechanically stable. Quasiparticle band structures reveal direct gaps in the near-infrared to visible range, with gap values increasing systematically from semilocal to hybrid exchange treatments. Optical spectra computed using a tight-binding Bethe-Salpeter approach demonstrate pronounced excitonic resonances arising from reduced dimensionality and weak dielectric screening. The exciton binding energies reach 152 meV in the 1T phase and 126 meV in the 2H phase, reflecting enhanced quantum confinement in the structurally denser phase. Our results identify \ce{Y2TeO2}monolayers as a rare class of stable, direct-gap MOenes with strong excitonic effects, providing a platform for exploring many-body physics in low-dimensional oxychalcogenide systems especially for photovoltaic applications. 

\end{abstract}


\section{Introduction}

The discovery of two-dimensional (2D) transition-metal carbides in 2011 marked a major milestone in materials science and led to the development of MXenes and related families such as MBenes \cite{vahidmohammadi2021world,Zhang2022_MBenes}. MXenes are typically produced by selectively removing the A-layer (e.g., Al, Si ) from layered ternary precursors known as MAX phases, such as Ti$_3$AlC$_2$, Ti$_3$SiC$_2$ and others \cite{anasori2019introduction}. This chemical exfoliation process yields atomically thin sheets like  Ti$_3$C$_2$ \cite{er2014ti3c2}. The general formula of MXenes is M$_{n+1}$X$_n$T$_x$ ($n = 1$, 2, 3 ...) \cite{anasori2019introduction}, where M is an early transition metal, X is carbon and/or nitrogen, and T$_x$ represents surface terminations introduced during synthesis. Several etching strategies have been developed to remove the A-layer and tailor the surface chemistry \cite{naguib2021ten}. Most conventional methods employ fluorine-containing etchants such as  HF~\cite{doi:10.1021/nn204153h}, LiF/HCl mixtures \cite{kewate2025role}, bifluoride salts~\cite{kiran2025comparative, ZHANG2022104384}, or fluorinated molten salts~\cite{https://doi.org/10.1002/bte2.20230021}. These approaches efficiently extract the A-layer and commonly result in --O, --OH, and --F surface groups~\cite{https://doi.org/10.1002/adma.202103148}. For instance, Kajiyama \textit{et al.}~\cite{kajiyama2017enhanced} demonstrated that LiF/HCl etching enhances electrochemical energy storage compared to direct HF treatment, partly due to increased interlayer spacing associated with --Cl species. However, --F, --O, and --OH groups remain dominant surface terminations.The presence and quantification of hydroxyl groups are still debated because hydrogen-related signals are difficult to detect experimentally~\cite{natu2023mxene}. To address excessive fluorine content, post-synthesis alkaline treatments using NaOH, TBAOH etc. have been proposed to partially replace halogen terminations with O/OH groups, thereby improving surface reactivity and chemical stability~\cite{natu2022effect,montazeri2023delamination}.

Surface terminations strongly influence the electronic properties of MXenes. Depending on composition and functionalization, MXenes can exhibit metallic, half-metallic, semiconducting, or even topologically nontrivial behavior~\cite{doi:10.1021/acs.nanolett.2c01914}. But, intrinsically semiconducting MXenes are relatively rare. Only a few systems, such as Mo$_2$CT$_x$ \cite{https://doi.org/10.1002/adfm.201505328}, and Mo$_2$TiC$_2$T$_x$~\cite{C5NH00125K}, have been realized experimentally and tested for various applications~\cite{Mahadi2025}. In the same line, many others have been predicted theoretically, including M$_2$CO$_2$ (M = Ti, Zr, Hf, Mn, W) \cite{gandi2016thermoelectric}, 2D Y$_2$CT$_x$I (T$_x$ = Br, Cl, F, H) \cite{aparicio2025theoretical}, MM'CT$_2$ (M = Sc; M' = Y; T = Br, Cl, F, H, I, O, OH, S, Se, Te) \cite{aparicio2025two}, Cr$_2$TiC$_2$T$_2$ (T = F, OH)~\cite{10.1063/1.4967983}, M$_2$CF$_2$ (M = Cr, Mo) \cite{zhou2024effect}, NbScCO$_2$ \cite{aparicio2025photovoltaic}, and M$_2$CT$_2$ (M = Sc, Y; T = O, F, OH)~\cite{aparicio2025solar}. The limited availability of stable semiconducting MXenes motivates the exploration of alternative 2D materials with tunable band gaps and robust structural stability.

In this direction, MXene-like metal-oxide monolayers, referred to as MOenes, have emerged as promising candidates \cite{Yan2024}. The prototype 1T-Ti$_2$O monolayer has been reported as a high-performance anode material and as a superconductor, while its fluorinated derivative 1T-Ti$_2$OF$_2$ exhibits a direct semiconducting band gap~\cite{doi:10.1021/acs.jpclett.0c03397}. Recent synthesis advances, including molten-salt electrochemical routes and thin-film growth on $\alpha$-Al$_2$O$_3$ substrates, indicate that MOenes are experimentally feasible and structurally stable~\cite{FAN2019607, D0TA01454K}. The oxide-based framework of MOenes also offers improved chemical robustness compared to carbide and nitride counterparts. Notably, MOenes show considerable potential across energy and electronic applications. For instance, Wan \textit{et al.}~\cite{10.1063/5.0196117} reported that Ti$_2$OX$_2$ (X = F, Cl) monolayers exhibit strong thermoelectric performance, combining high Seebeck coefficients with good electrical conductivity and low lattice thermal conductivity. Qiu \textit{et al.}~\cite{QIU2024136620} demonstrated that MOenes and their Janus structures possess enhanced piezoelectric responses compared to conventional 2D materials such as 2H-MoS$_2$, with Ti$_2$OH$_2$ showing an in-plane coefficient of  6.09~pm/V. In addition, Bolen \textit{et al.}~\cite{10.1063/5.0055701} found that Mo$_2$ScC$_2$O$_2$ combines a high Young’s modulus with low thermal conductivity, suggesting applications in mechanical sensing and thermal management.

Motivated by aforesaid developments, we investigate \ce{Y2TeO2} MOene monolayers in both 1T and 2H phases. Structural stability is evaluated using phonon dispersion, elastic constants, and ab initio molecular dynamics simulations. Electronic properties are examined through band structure and density of states analyses to determine band-gap characteristics and carrier behavior. Optical properties, including absorption coefficient, refractive index, and reflectivity, are calculated within the independent-particle framework and with electron–hole interactions included via the Bethe-Salpeter equation. Excitonic binding energies are extracted to assess many-body effects relevant for optoelectronic applications. Finally, the photovoltaic potential is estimated using the spectroscopic limited maximum efficiency and Shockley-Queisser approaches, yielding predicted power conversion efficiencies between 30.56\% and 32.66\%. These results highlight \ce{Y2TeO2} MOene as a promising 2D semiconductor for next-generation optoelectronic and energy-harvesting devices.

\section{Methodology}

First-principles simulations were performed within the framework of density functional theory (DFT)~\cite{Hohenberg1964, Kohn1965} using the Vienna \textit{Ab initio} Simulation Package (VASP, version~6.5.0)~\cite{Kresse_13115_1993, Kresse_11169_1996}. The exchange--correlation energy was treated using the Perdew--Burke--Ernzerhof (PBE) generalized gradient approximation~\cite{Perdew1996}, in combination with the projector augmented-wave (PAW) method~\cite{Blchl1994, Kresse1999}. A plane-wave kinetic energy cutoff of 700 eV was employed to ensure total-energy convergence.

Brillouin-zone integrations were carried out using Monkhorst--Pack \textbf{k}-point meshes of $12 \times 12 \times 1$ for structural relaxations and $36 \times 36 \times 1$ for electronic structure calculations. The electronic self-consistency criterion was set to 1e$^{-6}$ eV. Geometry optimizations were performed until the residual forces on each atom were smaller than 0.01 eV/\r{A}. A vacuum spacing of 23 \r{A} was introduced along the out-of-plane direction to eliminate spurious interactions between periodic images of the \ce{Y2TeO2}-1T and \ce{Y2TeO2}-2H monolayers. To obtain accurate band-gap values, electronic band structures were additionally calculated using the screened hybrid Heyd--Scuseria--Ernzerhof (HSE06) functional~\cite{Heyd2003, Krukau2006}.

We used the Phonopy package to perform phonon dispersion calculations to assess the system's dynamical stability~\cite {Togo_1_2015}. These computations were executed using density functional perturbation theory (DFPT) with a $2 \times 2 \times 1$ supercell and a $6 \times 6 \times 1$ \textbf{k}-point mesh. We used the Wannier90 code~\cite{Mostofi2008} to generate maximally localized Wannier functions (MLWFs) from a tight-binding (TB) Hamiltonian derived from HSE06 electronic structure calculations.

\begin{figure*}[!hbt]
    \centering
    \includegraphics[width=1.0\linewidth]{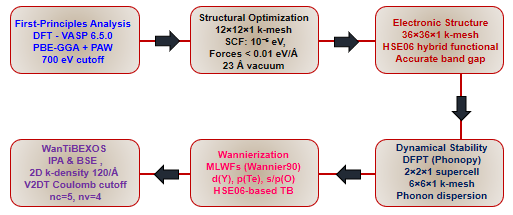}
    \caption{Block diagram steps and computational tools employed for the investigated 1T and 2H \ce{Y2TeO2} monolayers.}
    \label{fig:Block-diagram}
\end{figure*}

The WanTiBEXOS code~\cite{Dias2023} was used to figure out the optical and excitonic properties. The Wannier basis comprised $d$-orbital projections for Y atoms, $p$ orbitals for Te atoms, and $s$ and $p$ orbitals for O atoms. We examined optical responses in both the independent-particle approximation (IPA), which does not account for electron-hole interactions, and the Bethe-Salpeter equation (BSE) formalism, which does. For these calculations, a 2D \textbf{k}-point density of 120/\r{A} was used. To accurately characterize reduced-dimensional screening effects, BSE calculations employed a two-dimensional truncated Coulomb potential (V2DT)~\cite{Rozzi2006}. The dielectric function was calculated using a Gaussian smearing of 0.05 eV, with five conduction bands ($n_c = 5$) and four valence bands ($n_v = 4$) to capture low-energy optical and excitonic features accurately. A schematic representation of the parameters and computational tool simulation protocol is shown in Figure~\ref{fig:Block-diagram}.

\section{Results}

\subsection{Structural and mechanical properties}

\begin{figure*}[!hbt]
    \centering
    \includegraphics[width=1.0\linewidth]{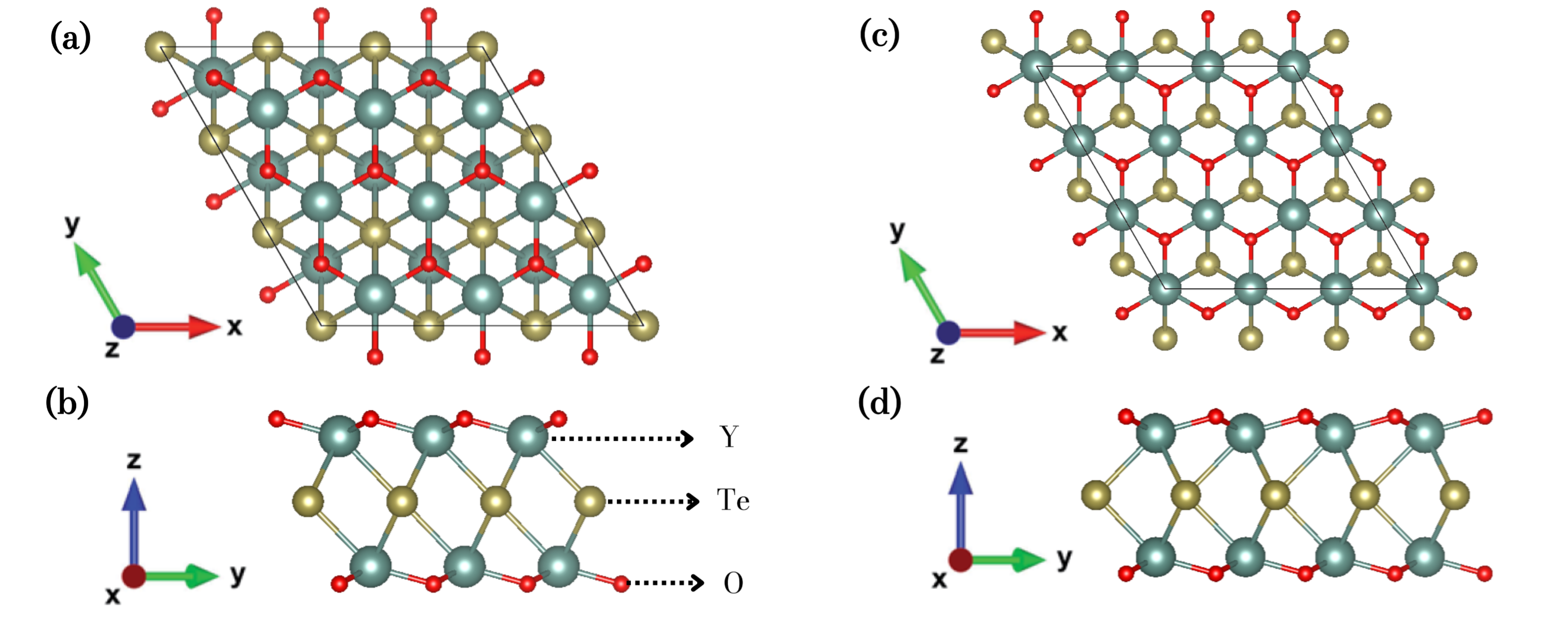}
    \caption{Structural models of the \ce{Y2TeO2} monolayers. Panels (a) and (b) show the top and side views of the \ce{Y2TeO2}-1T phase, while panels (c) and (d) display the corresponding views for the \ce{Y2TeO2}-2H phase. Y, Te, and O atoms are represented by golden-brown, silver-green, and red spheres, respectively.}
    \label{fig:structure}
\end{figure*}

Figure~\ref{fig:structure} shows the top and side views of the optimized atomic structures of the \ce{Y2TeO2} monolayers in the 1T and 2H phases. Both phases have a hexagonal lattice. The 1T phase belongs to the $P\bar{3}m1$ (No. 164) space group, and the 2H phase crystallizes in the $P\bar{6}m2$ (No. 187) symmetry. In both cases, the Y atoms create a close-packed hexagonal sublattice. The Te and O atoms, on the other hand, are in different coordination environments above and below the Y plane. This creates a layered metal-oxide framework that is typical of MOenes.

Table~\ref{Table:lattice} shows the lattice parameters, interaxial angles ($\alpha$, $\beta$, and $\gamma$), monolayer thicknesses, and cohesive energies that were found after complete structural relaxation at the PBE level. The in-plane lattice constants for the two phases are almost the same: $a_0 = 3.744$~\r{A} for \ce{Y2TeO2}-1T and $a_0 = 3.745$~\r{A} for \ce{Y2TeO2}-2H. This shows that changing the stacking order doesn't have a significant effect on the in-plane bonding network. These values align with earlier theoretical studies of analogous Y-based MOene systems, exhibiting deviations of < 1\% \cite{10.1063/1.4967983}, attributable to variations in exchange-correlation functionals and structural relaxation methodologies.

\begin{table}[!htb]
\begin{center}
\caption{Optimized structural parameters and energetic stability of the \ce{Y2TeO2} monolayers in the 1T and 2H phases. The table lists the in-plane lattice constant ($a_0$), interaxial angles ($\alpha$, $\beta$, and $\gamma$), monolayer thickness including a van der Waals length correction of 3.31~\r{A} ($t_0$), and the cohesive energy per atom ($E_{\text{coh}}$) calculated at the PBE level.}
\begin{tabular}{c c c c c}
\hline
 System & $a_0$ (\r{A}) & $\alpha=\beta\neq \gamma$ & $t_0$ (\r{A}) & $E_\text{coh}$ (eV/atom)  \\ \hline
 \ce{Y2TeO2}-1T & 3.744 & 90=90 $\neq$ 120 & 9.00 & -6.021 \\ 
 \ce{Y2TeO2}-2H & 3.745 & 90=90 $\neq$ 120 & 8.99 & -6.029 \\ 
\hline
\end{tabular}
\label{Table:lattice}
\end{center}
\end{table}

We calculated the cohesive energies of the proposed monolayers to assess their energetic stability.

\begin{equation} 
E_{\text{coh}} = \frac{E_{\text{tot}}(\ce{Y2TeO2}) - \sum_{i = Y, Te, O} N_i E_i}{N_{\text{atoms}}}, 
\end{equation}

\noindent where $E_{\text{tot}}(\ce{Y2TeO2})$ is the total energy of the monolayer, $N_i$ and $E_i$ are the number and isolated-atom energy of species $i$ (\ce{Y}, \ce{Te}, and \ce{O}), and $N_{\text{atoms}}$ is the total number of atoms in the unit cell. The cohesive energies of -6.021 eV/atom for the 1T phase and -6.029 eV/atom for the 2H phase show that the atoms are strongly bonded to each other and that the formation is thermodynamically favorable. These values are similar to those found for other stable 2D metal-oxide and MXene-derived systems~\cite{doi:10.1021/acs.jpclett.0c03397,FAN2019607}, which shows that both \ce{Y2TeO2} polymorphs are intrinsically stable in terms of energy.

\begin{figure}[!htb]
    \centering
    \includegraphics[width=0.8\linewidth]{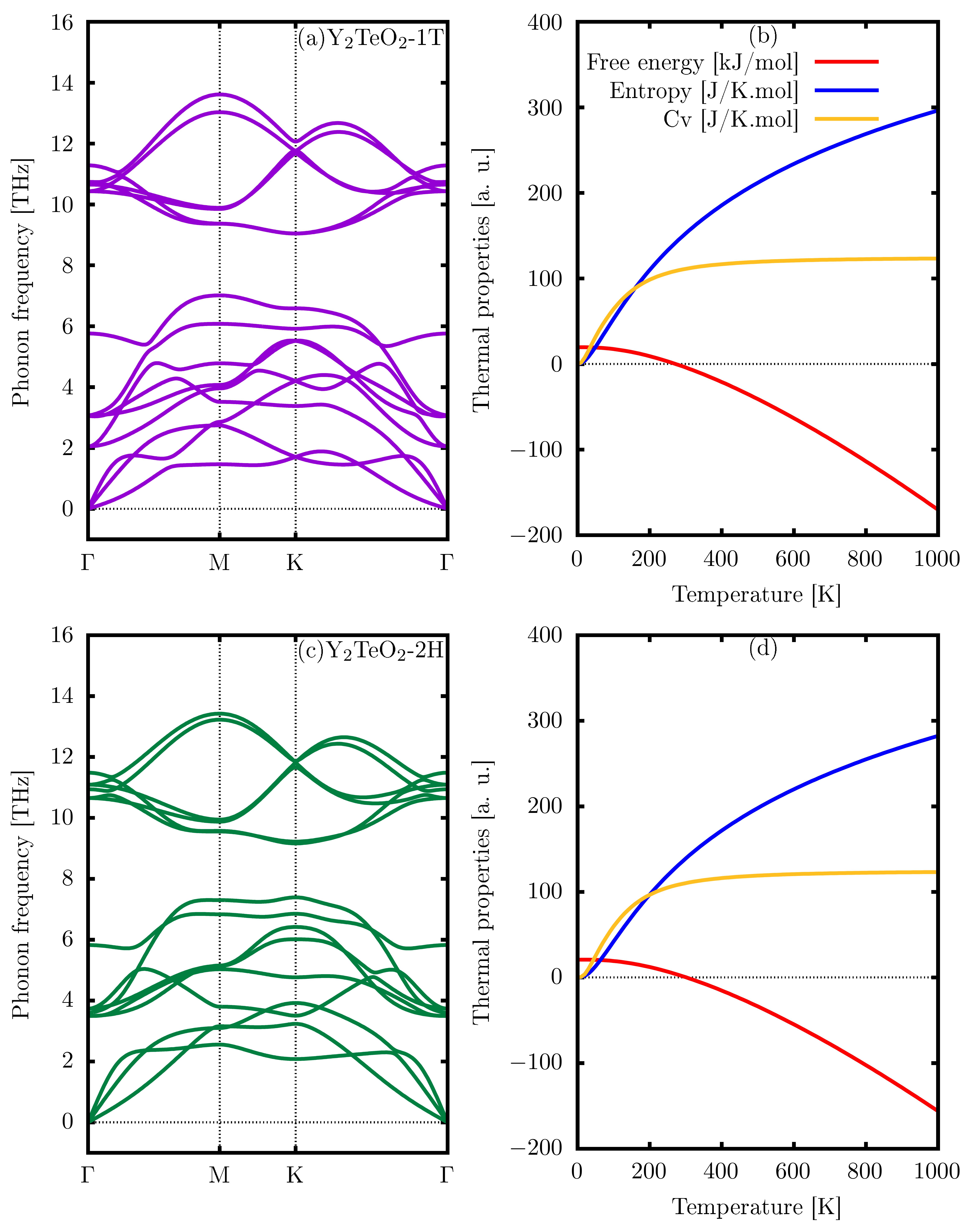}
    \caption{Phonon dispersion relations and thermodynamic properties of the \ce{Y2TeO2} monolayers. Panels (a) and (c) show the phonon dispersion curves of the \ce{Y2TeO2}-1T and \ce{Y2TeO2}-2H phases, respectively, calculated at the PBE level. Panels (b) and (d) display the temperature dependence of the Helmholtz free energy $F(T)$, entropy $S(T)$, and heat capacity at constant volume $C_v(T)$ for the corresponding monolayers.}
    \label{fig:phonons}
\end{figure}

We used phonon-dispersion calculations to assess the dynamic stability of \ce{Y2TeO2} monolayers in the 1T and 2H phases. You can see the results in Fig. \ref{fig:phonons}(a) and (c). In both instances, all phonon branches display positive frequencies across the entire Brillouin zone, with no imaginary modes observed. This behavior clearly indicates that both polymorphs are dynamically stable at zero temperature and that no structural instabilities arise from lattice distortions. The overall phonon spectrum shows that the acoustic and optical branches are well separated. This is a common feature in stable two-dimensional metal-oxide and MXene-derived systems. 

In addition to dynamic stability, the phonon density of states was used to find critical thermodynamic quantities like the Helmholtz free energy $F(T)$, entropy $S(T)$, and heat capacity at constant volume $C_v(T)$. These quantities are shown in Fig. ~\ref{fig:phonons}(b) and (d) for the 1T and 2H phases, respectively. The Helmholtz free energy decreases steadily as temperature increases, and it becomes negative at approximately $T \approx 350$ K for both monolayers. This trend indicates that thermal contributions help freestanding monolayers remain stable at moderate temperatures. This means they can be produced under conditions that can be tested in the laboratory.

The entropy $S(T)$ increases steadily with temperature. This is because phonon modes become more active, thereby increasing disorder in lattice vibrations. This behavior is consistent with observations in other stable two-dimensional materials, such as transition-metal dichalcogenides and oxide-based MXene analogues. It also supports the idea that the lattice dynamics in \ce{Y2TeO2} monolayers are strong.

The temperature dependence of the heat capacity, $C_v(T)$, provides additional information about how the system vibrates. At low temperatures ($T < 300$ K), $C_v$ exhibits a $T^3$ dependence consistent with the Debye model and the third law of thermodynamics, indicating that only long-wavelength acoustic phonons significantly influence the thermal response. At elevated temperatures ($T > 300$~K), $C_v$ progressively saturates, converging towards the classical Dulong–Petit limit ($C_v = 3R$)~\cite{mermin2006hans}. This saturation happens at about 100 J K$^{-1}$ mol$^{-1}$ for both the 1T and 2H phases. This means that all phonon modes are fully excited. The fact that the $C_v$ curves for the two polymorphs are so similar shows that changes in stacking sequence only have a small effect on their overall vibrational thermodynamics.

To further validate the structural stability of the \ce{Y2TeO2} monolayers, their mechanical stability was examined according to the Born–Huang criteria for hexagonal 2D systems~\cite{Haastrup2018}. These criteria say that the elastic constant $C_{11}$ must be positive ($C_{11} > 0$) and bigger than $C_{12}$ ($C_{11} > C_{12}$). Both the 1T and 2H phases meet these requirements. The in-plane elastic stiffness tensor $C_{ij}$ can be expressed as

\begin{equation}
C_{ij}=
\begin{pmatrix}
C_{11} & C_{12} & 0 \\
C_{12} & C_{11} & 0 \\
0 & 0 & C_{66} = \frac{C_{11}-C_{12}}{2}
\end{pmatrix},
\end{equation}

\noindent where symmetry relations impose $C_{11}=C_{22}$ and $C_{12}=C_{21}$. The calculated elastic constants for the \ce{Y2TeO2}-1T monolayer are $C_{11}=C_{22}=139.52$~N/m, $C_{12}=38.63$~N/m, and $C_{66}=51.68$~N/m. For the \ce{Y2TeO2}-2H phase, the corresponding values are $C_{11}=C_{22}=147.36$~N/m, $C_{12}=48.51$~N/m, and $C_{66}=41.42$~N/m. These values clearly fulfill the Born--Huang stability conditions~\cite{Haastrup2018}, confirming the mechanical robustness of both polymorphs.

\begin{figure*}[!htb]
    \centering
    \includegraphics[width=1.0\linewidth]{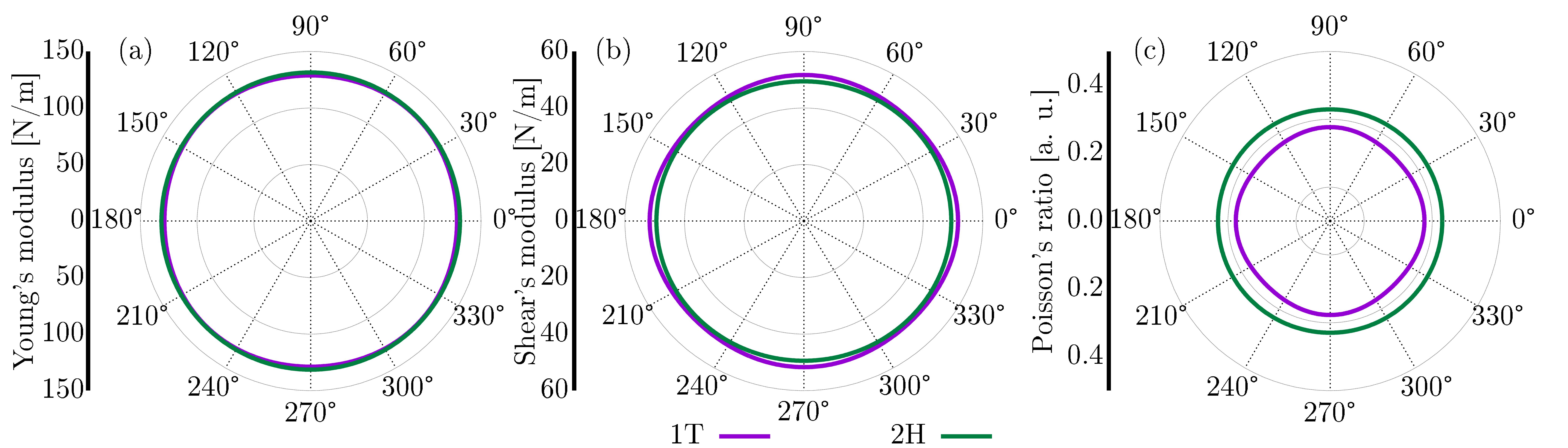}
    \caption{Polar plots of the in-plane mechanical properties of the \ce{Y2TeO2} monolayers: (a) Young’s modulus $Y(\theta)$, (b) shear modulus $G(\theta)$, and (c) Poisson’s ratio $\nu(\theta)$ for the 1T and 2H phases. The nearly isotropic angular dependence reflects the hexagonal symmetry and mechanical robustness of both structures.}
    \label{fig:elastic}
\end{figure*}

Using the elastic constants, we further investigated the directional dependence of the in-plane mechanical properties, namely the Young’s modulus $Y(\theta)$, shear modulus $G(\theta)$, and Poisson’s ratio $\nu(\theta)$, as functions of the in-plane angle $\theta \in [0,2\pi]$. The corresponding angular variations are presented in Fig.~\ref{fig:elastic}(a)--(c) and were obtained from Eqs.~(\ref{eq:y-theta})--(\ref{eq:g-theta})--(\ref{eq:v-theta}).

\begin{equation}
   Y(\theta)=\frac{C_{11}C_{22}-C^2_{12}}{C_{11}s^4+C_{22}c^4+\left( \frac{C_{11}C_{22}-C^2_{12}}{C_{66}}-2C_{12} \right)c^2s^2}, \label{eq:y-theta}
\end{equation}

\begin{equation}
   G(\theta) = \frac{C_{66}}{1+\left( C_{66}\frac{C_{11}+C_{22}+2C_{12}}{C_{11}C_{22}-C^2_{12}}-3 \right)c^2s^2} \label{eq:g-theta}
\end{equation}

\noindent and,

\begin{equation}
   \nu(\theta)=\frac{C_{12}(s^4+c^4)-\left(C_{11}+C_{12}-\frac{C_{11}C_{22}-C^2_{12}}{C_{66}} \right)c^2s^2}{C_{11}s^4+C_{22}c^4+\left( \frac{C_{11}C_{22}-C^2_{12}}{C_{66}}-2C_{12} \right)c^2s^2}, \label{eq:v-theta}
\end{equation}

\noindent where $s = \sin(\theta)$ and $c = \cos(\theta)$.

The polar plots reveal that $Y(\theta)$, $G(\theta)$, and $\nu(\theta)$ exhibit an almost isotropic angular dependence for both the 1T and 2H monolayers, consistent with their underlying hexagonal symmetry. Only minor deviations from perfect circularity are observed, which can be attributed to subtle differences in atomic stacking and bonding environments between the two phases.

Quantitatively, the maximum Young’s modulus of the \ce{Y2TeO2}-1T phase reaches $Y_{\text{max}}=130.39$~N/m, with a corresponding maximum shear modulus of $G_{\text{max}}=51.68$~N/m and a maximum Poisson’s ratio of $\nu_{\text{max}}=0.28$. For the \ce{Y2TeO2}-2H monolayer, the in-plane stiffness is slightly higher, with $Y_{\text{max}}=131.39$~N/m, while $G_{\text{max}}=49.43$~N/m and $\nu_{\text{max}}=0.33$. These values indicate that both polymorphs combine moderate in-plane stiffness with mechanical flexibility.

When compared to other two-dimensional materials, the Young’s moduli of \ce{Y2TeO2} monolayers are lower than that of graphene ($\sim$340~N/m), but are comparable to or exceed those of many transition-metal dichalcogenides, such as MoS$_2$ ($\sim$120~N/m), and several functionalized MXenes. This balance between stiffness and isotropy is particularly advantageous for applications in flexible optoelectronic and photovoltaic devices, where mechanical integrity under strain is a critical requirement.

\subsection{Electronic properties}

\begin{figure}[!htb]
    \centering
    \includegraphics[width=0.7\linewidth]{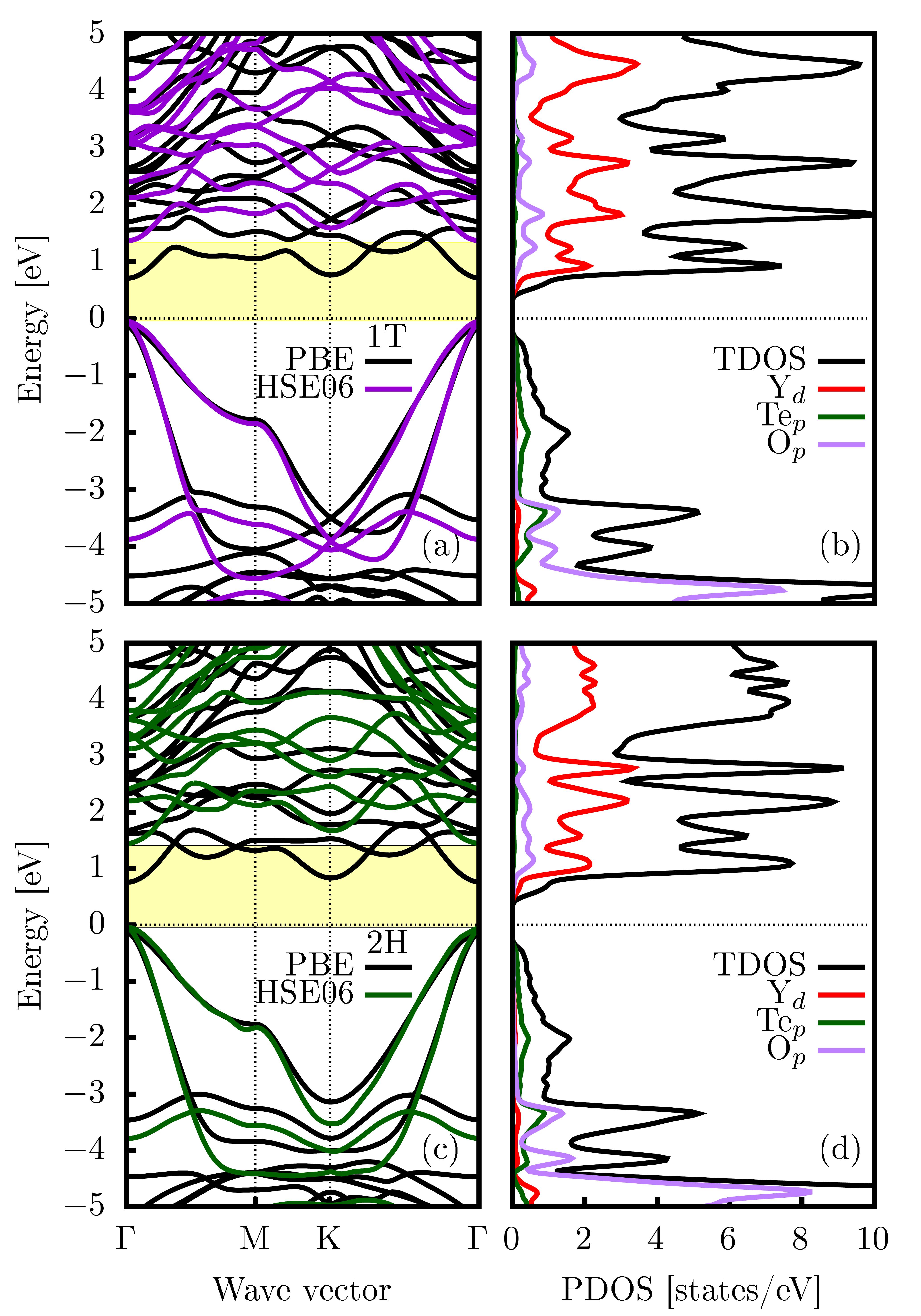}
    \caption{Electronic band structures and density of states of the \ce{Y2TeO2} monolayers. Panels (a) and (c) show the electronic band structures of the \ce{Y2TeO2}-1T and \ce{Y2TeO2}-2H phases, respectively, calculated at the PBE level (black lines) and with the HSE06 hybrid functional (colored lines). Panels (b) and (d) display the corresponding total density of states (TDOS) and projected density of states (PDOS), decomposed into contributions from Y $d$, Te $p$, and O $p$ orbitals. The Fermi level is set to zero energy.}
    \label{fig:bandas}
\end{figure}

Figure~\ref{fig:bandas}(a) and (c) show the electronic band structures of the \ce{Y2TeO2} monolayers in the 1T and 2H phases, respectively. At the PBE level, both monolayers have a direct band gap at the $\Gamma$ point. The gap value for \ce{Y2TeO2}-1T is 0.84 eV and the gap value for \ce{Y2TeO2}-2H is 0.89 eV. The direct nature of the gap is especially useful for optoelectronic applications because it makes light absorption and radiative recombination more efficient without the need for phonon-assisted transitions.

It is well known that semilocal exchange--correlation functionals such as PBE tend to underestimate band gaps due to self-interaction errors and the absence of nonlocal exchange effects, which result in an overdelocalization of electronic states. To obtain more reliable electronic gaps, we therefore employed the screened hybrid HSE06 functional. As shown in Fig.~\ref{fig:bandas}, the inclusion of a fraction of exact Hartree--Fock exchange leads to a significant opening of the band gap, yielding values of 1.42~eV for the 1T phase and 1.47~eV for the 2H phase, while preserving the direct-gap character at the $\Gamma$ point. These HSE06 band gaps fall within the optimal range for single-junction photovoltaic absorbers according to the Shockley--Queisser criterion and are comparable to those of widely studied two-dimensional semiconductors, such as monolayer MoS$_2$ (1.6--1.9~eV) and WS$_2$  \cite{Luan2017_WS2_MoS2,Joseph2021}. Compared to most functionalized MXenes, which are typically metallic or exhibit very small gaps, the intrinsic semiconducting behavior of \ce{Y2TeO2} monolayers represents a significant advantage for optoelectronic and energy-harvesting applications.

Figure~\ref{fig:bandas}(b) and (d) show the total and projected density of states (TDOS and PDOS) for the \ce{Y2TeO2}-1T and \ce{Y2TeO2}-2H monolayers. The PDOS analysis shows that the conduction band minimum is primarily composed of Y $d$ states, with a small contribution from O $p$ orbitals. This indicates that metal and oxygen states are mixing. The valence band maximum, by contrast, arises primarily from O $p$ orbitals, with smaller contributions from Y $d$ and Te $p$ states. This orbital distribution clarifies the origin of the direct band gap and highlights the metal--oxygen framework as the key factor governing the electronic structure of \ce{Y2TeO2} MOenes. The dominant Y $d$ character near the conduction band edge suggests relatively low electron effective masses, which is beneficial for charge transport, while the O $p$-dominated valence band is expected to support strong light--matter interactions. Similar orbital features have been reported for other metal--oxide-based two-dimensional systems, further reinforcing the consistency of our results with the broader class of MOene materials.

\subsection{Excitonic and optical properties}

\begin{figure}[!htb]
    \centering
    \includegraphics[width=0.5\linewidth]{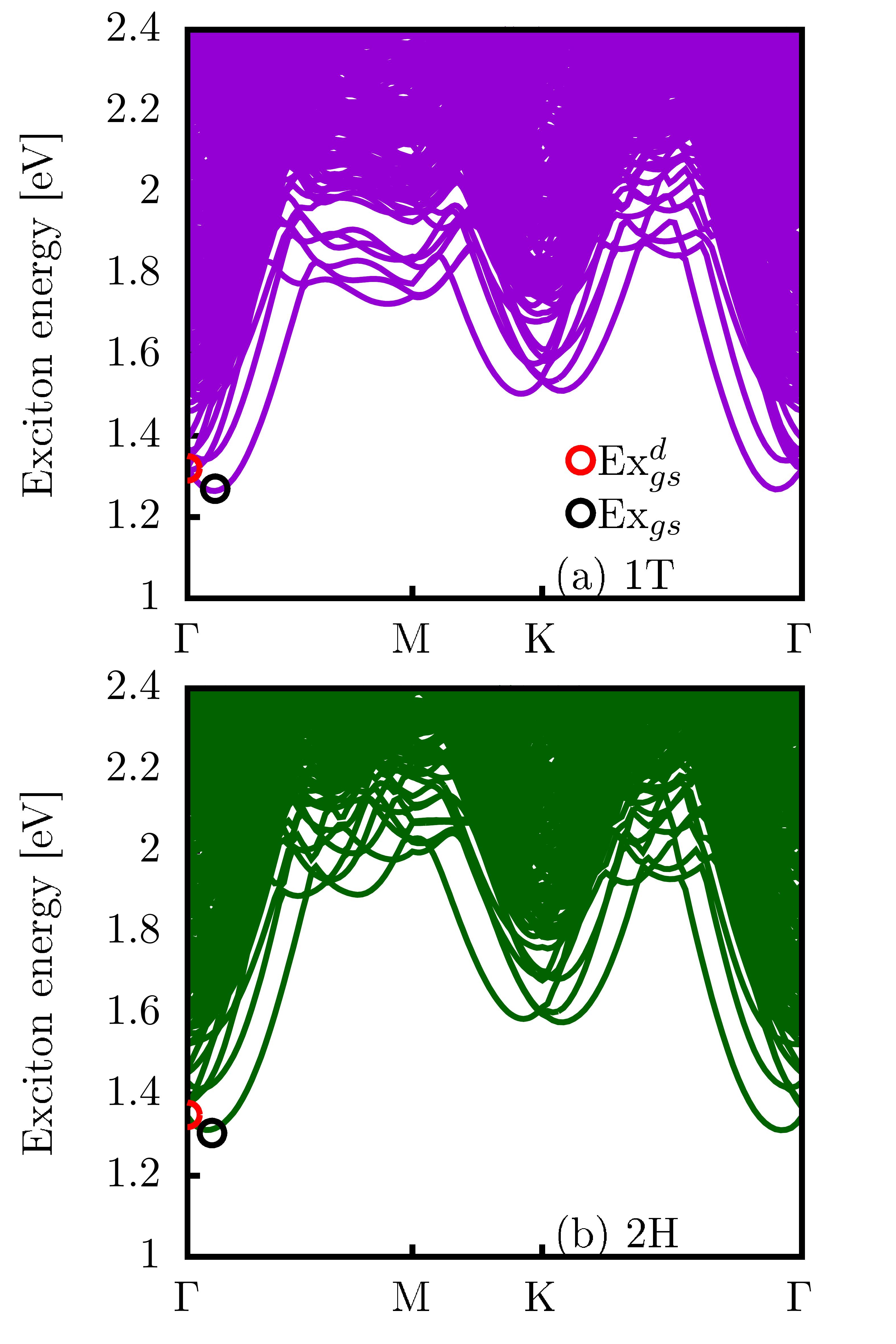}
    \caption{Exciton band structures of the \ce{Y2TeO2} monolayers obtained from the MLWF-TB+BSE approach. Panels (a) and (b) correspond to the \ce{Y2TeO2}-1T and \ce{Y2TeO2}-2H phases, respectively. The lowest-energy indirect exciton ground states ($\mathrm{Ex}_{gs}$) and the direct excitonic states at the $\Gamma$ point ($\mathrm{Ex}^{d}_{gs}$) are highlighted.}
    \label{fig:excitons}
\end{figure}

The excitonic properties of the \ce{Y2TeO2} monolayers were investigated by analyzing the exciton band structures obtained within the MLWF-TB+BSE framework, as shown in Fig.~\ref{fig:excitons}(a) and (b) for the 1T and 2H phases, respectively. Although both monolayers exhibit a direct electronic band gap at the $\Gamma$ point, their exciton ground states are found to be indirect in momentum space. Specifically, the lowest-energy excitons are located near the $\Gamma$--M region, with excitation energies of 1.26~eV for \ce{Y2TeO2}-1T and 1.31~eV for \ce{Y2TeO2}-2H.

The exciton ground state is indirect because the excitonic bands spread out, which is a result of the interaction between the electronic band structure and electron-hole interactions. This behavior has also been observed in other 2D semiconductors \cite{Aparicio_2026_Sc2CandY2C,Qiu2013}, where excitonic effects can alter the optical response in relation to the underlying electronic gap. As a result, phonon-assisted optical transitions with energies below the direct optical gap can occur, which may affect the absorption onset and radiative recombination processes.

The exciton binding energies, which are shown in Table~\ref{Table-excitons}, were found by taking the difference between the fundamental electronic band gap and the exciton ground-state energy. The binding energies calculated are 152 meV for the 1T phase and 126 meV for the 2H phase, which means that the excitons are moderately bound. These numbers are much higher than what is usually seen in bulk semiconductors, but they are lower than the binding energies that have been seen in strongly confined two-dimensional systems like monolayer transition-metal dichalcogenides, where values over 300 meV are common. This intermediate regime shows a balance between lower dimensionality and better dielectric screening in \ce{Y2TeO2} MOenes.

\begin{table}[!htb]
\begin{center}
\caption{Excitonic properties of the \ce{Y2TeO2} monolayers obtained using the MLWF-TB+BSE method with HSE06 parametrization. The table lists the fundamental electronic band gap ($E_g$), direct band gap ($E_g^{d}$), exciton ground-state energy ($\mathrm{Ex}_{gs}$), direct exciton ground-state energy ($\mathrm{Ex}^{d}_{gs}$), and the exciton binding energy ($\mathrm{Ex}_b = E_g - \mathrm{Ex}_{gs}$) for the 1T and 2H phases.}
\vspace{0.2 cm}
\begin{tabular}{l c c}
\toprule
Parameter & \ce{Y2TeO2}-1T & \ce{Y2TeO2}-2H \\
\midrule
$E_g$ PBE  (eV)           & 0.84 & 0.89 \\
$E_g$ HSE06  (eV)         & 1.42 & 1.47 \\
$E^{d}_g$    (eV)         & 1.42 & 1.47 \\
$\text{Ex}_{gs}$ (eV)     & 1.26 & 1.31 \\
$\text{Ex}^{d}_{gs}$ (eV) & 1.32 & 1.35 \\
$\text{Ex}_b$ (meV)  & 152  & 126 \\
\bottomrule
\end{tabular}
\label{Table-excitons}
\end{center}
\end{table}

We found direct excitonic transitions at the $\Gamma$ point, with energies of 1.32 eV for \ce{Y2TeO2}-1T and 1.35 eV for \ce{Y2TeO2}-2H. The direct exciton binding energies are equal to the indirect ground state binding energies. This further demonstrates that the interactions between electrons and holes are the primary influence on the low-energy optical response. The fact that the direct exciton energies are close to the best spectral range for absorbing visible and near-infrared light shows how useful \ce{Y2TeO2} monolayers are for optoelectronic and photovoltaic applications.

\begin{figure*}[!htb]
    \centering
    \includegraphics[width=0.7\linewidth]{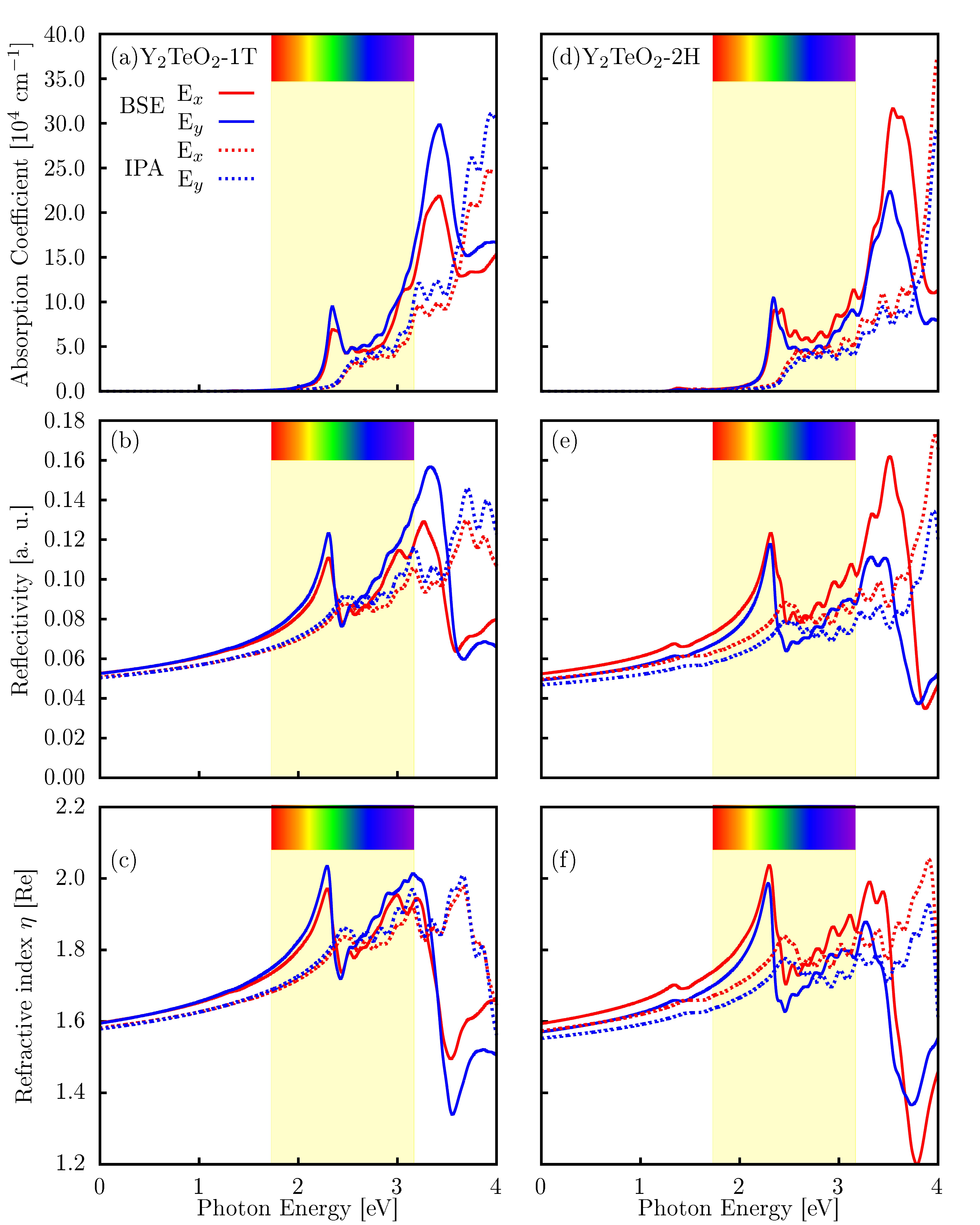}
    \caption{Optical properties of the \ce{Y2TeO2} monolayers calculated within the independent-particle approximation (IPA) and the Bethe--Salpeter equation (BSE) framework. Panels (a) and (d) show the absorption coefficient $\alpha(\omega)$, panels (b) and (e) the reflectivity $R(\omega)$, and panels (c) and (f) the refractive index $\eta(\omega)$ for the \ce{Y2TeO2}-1T and \ce{Y2TeO2}-2H phases, respectively. Solid and dashed curves correspond to BSE and IPA results, while blue and red lines denote light polarized along the $\hat{x}$ and $\hat{y}$ directions. The shaded region highlights the visible spectral range.}
    \label{fig:optical}
\end{figure*}

Figure~\ref{fig:optical} shows the optical response of the \ce{Y2TeO2}-1T and \ce{Y2TeO2}-2H monolayers. This includes the absorption coefficient $\alpha(\omega)$, reflectivity $R(\omega)$, and refractive index $\eta(\omega)$, which were calculated using both the independent-particle approximation (IPA) and the Bethe–Salpeter equation (BSE) formalism. The optical spectra were analyzed for in-plane light polarizations along the $\hat{x}$ and $\hat{y}$ directions, facilitating the evaluation of potential in-plane anisotropy.

Figure~\ref{fig:optical}(a) and (d) show the absorption spectra. They show that both monolayers absorb a lot of light in the visible and ultraviolet (UV) regions. The \ce{Y2TeO2}-2H phase shows an almost isotropic absorption response at the IPA level (dashed curves), while the 1T phase shows a polarization dependence that is a little stronger. When the BSE formalism is used to explicitly include excitonic effects (solid curves), both polymorphs show a systematic red shift of the absorption onset. This is because bound electron-hole pairs are forming. This red shift caused by excitons is a typical property of two-dimensional semiconductors. It shows how important many-body effects are for accurately describing their optical properties.

Even though both structures have a hexagonal symmetry, the presence of excitonic effects causes a weak optical anisotropy, as shown by the small differences between the $\hat{x}$ and $\hat{y}$ polarized responses. The absorption spectra in the two in-plane directions are very similar, which means that both \ce{Y2TeO2} monolayers can be thought of as quasi-isotropic optical absorbers. The existence of significant absorption peaks in the visible spectrum reinforces their suitability for optoelectronic and photovoltaic applications.

The reflectivity spectra shown in Fig. ~\ref{fig:optical}(b) and (e) show the same patterns at both the IPA and BSE levels. The reflectivity generally rises with the energy of the photons, peaking in the UV range. At the BSE level, the \ce{Y2TeO2}-1T monolayer has a maximum reflectivity of about 15.8\%, and at the IPA level, it has a maximum reflectivity of about 14.3\%. The \ce{Y2TeO2}-2H phase, on the other hand, has a maximum reflectivity of about 16.2\% (BSE) and 17.0\% (IPA) for $\hat{x}$ polarization. These moderate reflectivity values are good for light-harvesting applications because they mean that less light is lost when it reflects.

Figures~\ref{fig:optical}(c) and (f) show the refractive index $\eta(\omega)$ that was found using the IPA and BSE methods. The main peaks in the refractive index for the \ce{Y2TeO2}-1T monolayer are in the visible and near-UV ranges, around 2.42 eV and 3.16 eV at the BSE level. In the $\hat{x}$ direction, $\eta$ ranges from 1.35 to 2.15 (1.55 to 2.00 at the IPA level). In the $\hat{y}$ direction, it ranges from 1.50 to 1.95 (1.55 to 2.02 at the IPA level). The refractive index for the \ce{Y2TeO2}-2H monolayer is between 1.20 and 2.05 (1.58 and 2.06 at the IPA level) along $\hat{x}$ and between 1.37 and 1.98 (1.56 and 1.90 at the IPA level) along $\hat{y}$.

\subsection{Power conversion efficiency (PCE)}

The optical properties discussed above show that \ce{Y2TeO2} monolayers have strong visible-light absorption and band gaps that are close to the best range for photovoltaic applications. To quantitatively evaluate their efficacy as solar absorbers, we calculated the power conversion efficiency (PCE) employing both the spectroscopic limited maximum efficiency (SLME) framework~\cite{Yu2012} and the SQ detailed-balance limit~\cite{Shockley1961}. These methods enable a coherent assessment of the interactions among photon absorption, radiative recombination, and electronic band-gap properties, accounting for optical responses at both the independent-particle approximation (IPA) and the Bethe–Salpeter equation (BSE) levels.

The PCE is defined as
\begin{equation}
\mathrm{PCE} = \frac{J_{\mathrm{SC}} V_{\mathrm{OC}} \mathrm{FF}}{P_{\mathrm{SOLAR}}},
\end{equation}
\noindent where $J_{\mathrm{SC}}$ is the short-circuit current density, $V_{\mathrm{OC}}$ the open-circuit voltage, FF the fill factor, and $P_{\mathrm{SOLAR}}$ the total incident solar power per unit area. The short-circuit current density is obtained from
\begin{equation}
J_{\mathrm{SC}} = \int_{E_g^{d}}^{\infty} \frac{P(E)}{E} \, \mathrm{d}E,
\end{equation}
\noindent while the total solar power is given by
\begin{equation}
P_{\mathrm{SOLAR}} = \int_{0}^{\infty} P(E)\, \mathrm{d}E,
\end{equation}
\noindent with $P(E)$ representing the AM1.5G solar spectral irradiance~\cite{Dias2023, Astm_1_2012}.

The current--voltage characteristics of the illuminated solar cell determine the output power density, whose maximum value,
\begin{equation}
P_{\mathrm{PV}} = J(V_{\mathrm{max}}) V_{\mathrm{max}},
\end{equation}
\noindent is obtained at the operating voltage $V_{\mathrm{max}}$. The current density $J(V)$ is expressed as
\begin{equation}
J(V) = J_{\mathrm{SC}} - \frac{J_0}{f_r} \left[ \exp\!\left(\frac{eV}{k_{\mathrm{B}}T}\right) - 1 \right],
\end{equation}
\noindent where $J_0$ is the reverse saturation current density, $f_r$ the radiative recombination fraction, $e$ the elementary charge, $k_{\mathrm{B}}$ the Boltzmann constant, and $T$ the device temperature.

Within the SLME framework, the photogenerated current density is evaluated as
\begin{equation}
J_{\mathrm{SC}} = e \int_{0}^{\infty} a(E)\frac{P(E)}{E}\,\mathrm{d}E,
\end{equation}
\noindent where $a(E)$ is the absorbance. The reverse saturation current density is determined from detailed balance under thermal equilibrium,
\begin{equation}
J_0 = e\pi \int_{0}^{\infty} a(E)\Phi_{\mathrm{bb}}(E)\,\mathrm{d}E,
\end{equation}
\noindent with the blackbody photon flux given by
\begin{equation}
\Phi_{\mathrm{bb}}(E) = \frac{2E^2}{h^3 c^2} \left[\exp\!\left(\frac{E}{k_{\mathrm{B}}T}\right)-1\right]^{-1}.
\end{equation}

In Table~\ref{Table-pces}, you can see a summary of the PCE values we got for the \ce{Y2TeO2} monolayers in the 1T and 2H phases at both the IPA and BSE optical levels. The SLME efficiencies calculated using the realistic absorbance of freestanding monolayers are under 0.3\%. This is because two-dimensional materials are only a few atoms thick, which makes it very hard for light to be absorbed. Other pristine 2D absorbers have also been found to have low efficiencies, underscoring the difficulty of using isolated monolayers directly in photovoltaic devices.

\begin{table}[!htb]
\centering
\caption{Calculated power conversion efficiencies (PCEs) of the \ce{Y2TeO2} monolayers in the 1T and 2H phases obtained within the SLME and SQ frameworks at $T = 300$~K compared to literature. PCE$^{\mathrm{SLME}}$ corresponds to efficiencies calculated using the realistic absorbance of freestanding monolayers at the IPA and BSE optical levels. PCE$_{\mathrm{max}}^{\mathrm{SLME}}$ denotes the maximum SLME assuming ideal absorption above the direct band gap, while PCE$^{\mathrm{SQ}}$ represents the SQ detailed-balance limit.}
\begin{tabular}{c c c c c c}
\toprule
System & Level & PCE$^{\text{SLME}}$  & PCE$_\text{max}^{\text{SLME}}$  & PCE$^{\text{SQ}}$ & Ref.\\
\midrule
\ce{Y2TeO2}-1T & IPA & 0.16 & 32.27 & 32.27 & This work  \\
\ce{Y2TeO2}-1T & BSE & 0.28 & 30.56 & 32.35 & This work  \\ 
\ce{Y2TeO2}-2H & IPA & 0.15 & 31.55 & 31.49 & This work  \\
\ce{Y2TeO2}-2H & BSE & 0.29 & 31.57 & 32.66 & This work  \\
\ce{ScNbCO2}-1T & IPA & - & 14.33 & 20.20 & Ref.\cite{aparicio2025photovoltaic}  \\
\ce{ScNbCO2}-1T & BSE & - & 20.68 & 29.27 &   \\ 
\ce{ScYC(OH)2}-1T & IPA & 0.45 & 29.76 & 29.76 & Ref.\cite{aparicio2025two}  \\
\ce{ScYC(OH)2}-1T & BSE & 0.32 & 16.82 & 23.44 &   \\ 
\ce{Y2CCl2}-1T & IPA & 0.93 & 26.31 & 27.32 & Ref.\cite{Aparicio2025enhanced}  \\
\ce{Y2CCl2}-1T & BSE & 0.99 & 32.67 & 32.67 &   \\ 
\ce{Y2CBr2}-1T & IPA & 0.61 & 14.10 & 19.32 & Ref.\cite{Aparicio2025enhanced}  \\
\ce{Y2CBr2}-1T & BSE & 1.02 & 31.73 & 32.08 &   \\
\ce{Y2CI2}-1T & IPA & 0.95 & 19.62 & 29.24 & Ref.\cite{Aparicio2025enhanced}  \\
\ce{Y2CI2}-1T & BSE & 1.23 & 29.85 & 32.12 &   \\
\ce{Y2CH2}-1T & IPA & 0.85 & 27.14 & 28.47 & Ref.\cite{Aparicio2025enhanced}  \\
\ce{Y2CH2}-1T & BSE & 0.79 & 29.85 & 32.11 &   \\
\ce{SiI2}-1T & IPA & 0.24 & 6.45  & 7.84 & Ref.\cite{Aparicio2025_XI2}  \\
\ce{SiI2}-1T & BSE & 0.54 & 13.57 & 16.37 &   \\
\ce{PdS2}-1T & IPA & 1.03 & 16.64  & 18.26 & Ref.\cite{Moujaes2023}  \\
\ce{PdS2}-1T & BSE & 1.28 & 27.89 & 28.95 &   \\
\ce{PdSe2}-1T & IPA & 1.26 & 21.19  & 29.88 & Ref.\cite{Moujaes2023}  \\
\ce{PdSe2}-1T & BSE & 1.24 & 26.45 & 32.32 &   \\
\bottomrule
\end{tabular}
\label{Table-pces}
\end{table}

Prior research has shown that light-trapping techniques, optical cavities, and multilayer stacking can significantly increase the effective absorbance of two-dimensional materials, potentially nearing unity~\cite{Jariwala2017}. These results led us to look into the highest possible SLME ($\mathrm{PCE}_{\mathrm{max}}^{\mathrm{SLME}}$), assuming that all photons with energies above the direct band gap are absorbed perfectly. In this idealized situation, the \ce{Y2TeO2}-1T monolayer has a maximum PCE of 32.27\% (IPA) and 30.56\% (BSE), while the \ce{Y2TeO2}-2H phase has a maximum PCE of 31.55\% (IPA) and 31.57\% (BSE). These numbers are close to the Shockley–Queisser limit, indicating that \ce{Y2TeO2} has favorable band-gap alignment for solar energy conversion.

The SQ analysis predicts that single-junction solar cells have the optimal band gap of approximately 1.4 eV~\cite{Shockley1961}. The direct HSE06 band gaps of 1.42 eV for the 1T phase and 1.47 eV for the 2H phase are nearly identical—the SQ efficiencies at the IPA level range from 31.49\% to 32.27\%. When excitonic effects are included using the BSE formalism, the values increase slightly to 32.35\% and 32.66\%. This improvement demonstrates how excitonic interactions can alter the optical absorption profile and increase the theoretical upper limits of photovoltaic performance.

Table \ref{Table-pces} also presents PCEs of different materials predicted at SLME and SQ levels, which are carefully listed. The proposed \ce{Y2TeO2} monolayers have an excellent PCE excitonic solar cells compared to the literature . For example, the \ce{PCE^{SQ}} of the \ce{ScNbCO2}-1T phase monolayer at the IPA (BSE) levels are 20.20\% (29.27\%) \cite{aparicio2025photovoltaic}. Likewise, \ce{ScYC(OH)2}-1T phase showed \ce{PCE^{SQ}} values of 29.76\% (23.44\%) \cite{aparicio2025two}. Yttrium MXenes monolayers were reported with values of 19.32 \% to 29.24\% (32.11 \% to32.67 \%) at the IPA (BSE) levels \cite{Aparicio2025enhanced}, respectively. In addition, silicon diiodide \ce{XI2}-1T was reported with \ce{PCE^{SQ}} of 7.84\% (16.37\%) at the IPA (BSE) levels \cite{Aparicio2025_XI2}. Also, \ce{PdS2}-1T and \ce{PdSe2}-1T monolayers showed  values of  18.26\% and  29.88\%  at IPA level, while 28.95\% and  32.32\% at the BSE level \cite{Moujaes2023}. Our optimal results suggest that as long as the material has a band gap value between 1.30 eV and 1.40 eV, it will exhibit an excellent response because this is an optimal solar region \cite{Shockley1961,Jariwala2017}.

It is important to mention that a band gap of 1.42 eV for the 1T phase and 1.47 for the 2H phase is close to the optimum region for solar harvesting. In addition, a good PCE means that the materials have a low exciton binding energy ranging between 126 meV (\ce{Y2TeO2}-2H) and 152 meV (\ce{Y2TeO2}-1T), which would facilitate efficient exciton dissociation into free charge carriers \cite{Huacarpuma2026,Massicotte2018}. This suggests that \ce{Y2TeO2} monolayers in IT and 2H phases achive efficiencies comparable to or higher than those of other recently proposed 2D monolayers.

\section{Conclusions}

Herein, we present a systematic first-principles investigation of the structural stability and optoelectronic performance of \ce{Y2TeO2} MOene monolayers in the 1T and 2H phases. Both polymorphs satisfy the criteria for dynamical and mechanical stability, as confirmed by phonon dispersion, elastic analysis, and exhibit nearly isotropic in-plane elastic behavior with stiffness comparable to other robust 2D nanosystems. These features indicate that \ce{Y2TeO2} can sustain mechanical integrity under practical device conditions. Electronic structure calculations reveal direct band gaps at the $\Gamma$ point of 1.42 eV (1T) and 1.47 eV (2H), placing both phases within the optimal range for single-junction photovoltaic absorbers. The similarity in band-gap magnitude across phases suggests structural flexibility without significant compromise in electronic performance.

Many-body analysis further shows that electron-hole interactions lead to moderately bound excitons, with binding energies of 152 meV and 126 meV for the 1T and 2H phases, respectively. These values indicate appreciable Coulomb interaction in the 2D limit while remaining favorable for exciton dissociation under operational conditions. The inclusion of excitonic effects produces a red shift in the optical absorption edge, highlighting the importance of beyond-independent-particle treatments for accurate optical prediction. The optical response demonstrates strong absorption across the visible and ultraviolet regions, moderate reflectivity, and weak in-plane anisotropy, supporting broadband light-harvesting capability. Although an isolated monolayer inherently limits absolute light absorption due to atomic thickness, the calculated spectroscopic limited maximum efficiency and Shockley-Queisser limits reach ~ 33\%. This high theoretical efficiency originates from the near-ideal band-gap alignment and favorable absorption characteristics. Thus, our results establish \ce{Y2TeO2} MOene as a structurally stable, direct-band-gap 2D semiconductor with balanced excitonic and optical properties. When integrated into multilayer stacks, van der Waals heterostructures, and light-management architectures, these monolayers hold strong potential for next-generation optoelectronic and photovoltaic applications.

\section*{Acknowledgments}
This work was supported by the Brazilian funding agencies Fundação de Amparo à Pesquisa do Estado de São Paulo--FAPESP, grants 2024/21870-8, 2022/16509-9, and 2024/05087-1), and National Council for Scientific, Technological Development--CNPq, grant no. 307213/2021–8.The authors gratefully acknowledge financial support from the following institutions: the National Council for Scientific and Technological Development (CNPq), the Federal District Research Support Foundation (FAPDF), and the Coordination for Improvement of Higher Education Personnel (CAPES). The authors also express their gratitude to the National Laboratory for Scientific Computing for providing resources through the Santos Dumont supercomputer, and to the ``Centro Nacional de Processamento de Alto Desempenho em S\~ao Paulo'' (CENAPAD-SP, UNICAMP/FINEP - MCTI project) for support related to projects 897 and 909. Additional resources were provided by Lobo Carneiro HPC (NACAD) at the Federal University of Rio de Janeiro (UFRJ) for project 133. L.A.R.J acknowledges financial support from FAPDF-PRONEM grant 00193.00001247 /2021-20, PDPG-FAPDF-CAPES Centro-Oeste grant number 00193-00000867/2024-94, and CNPq grants 301577/2025-0 and 444111/2024-7. A.C.D. acknowledges financial support from FAP-DF grants 00193-00001817/2023-43 and 00193-00002073/2023-84, CNPq grants 408144/2022-0, 305174/2023-1, 444069/2024-0, and 444431/2024-1. 

\bibliographystyle{unsrtnat}
\bibliography{references}  

\end{document}